\documentclass[11pt]{article}
\usepackage[dvips]{graphicx}
\usepackage{a4wide}

\newcommand{\nn}{\nonumber}
\newcommand{\be}{\begin{equation}}
\newcommand{\ee}{\end{equation}}
\newcommand{\ba}{\begin{eqnarray}}
\newcommand{\ea}{\end{eqnarray}}
\newcommand{\ci}[1]{\cite{#1}}

\def\vd{{\bf \Delta}_\perp}
\def\vo{{\bf 0}_\perp}
\def\vea{{\bf e}_1}
\def\veb{{\bf e}_2}

\def\als{\alpha_s}

\def\mev{\,{\rm MeV}}
\def\gev{\,{\rm GeV}}

\newcommand{\sla}{\hspace*{-0.20cm}/}

\newcommand{\req}[1]{(\ref{#1})}
\def\xb{x}
\def\ub{\overline{u}}

\def\veps{\varepsilon}
\def\sh{\hat{s}}
\def\uh{\hat{u}}
\def\th{\hat{t}}
\def\kb{\overline{k}}
\def\zb{\overline{z}}
\def\kpt{\kappa_{\scriptscriptstyle T}^{\scriptscriptstyle P}}

\begin{document}
\thispagestyle{empty}
\begin{flushright}
WU B 03-05 \\
hep-ph/0309071\\
September 2003\\[20mm]
\end{flushright}

\begin{center}
{\Large\bf Signatures of the handbag mechanism in \\[0.3em]
 wide-angle photoproduction \\[0.4em]
 of pseudoscalar mesons} \\
\vskip 15mm

H.W.\ Huang \\[1em]
{\small {\it Department of Physics, University of Colorado, Boulder, 
CO 80309-0390, USA}}\\
\vskip 5mm

R. Jakob and P.\ Kroll \\[1em]
{\small {\it Fachbereich Physik, Universit\"at Wuppertal, D-42097 Wuppertal,
Germany}}\\
\vskip 5mm

K. Passek-Kumeri\v{c}ki\\[1em]
{\small {\it Theoretical Physics Division, Rudjer Bo\v{s}kovi\'{c} Institute, 
HR-10002 Zagreb, Croatia}}\\
\end{center}

\begin{abstract}
Wide-angle photoproduction of pseudoscalar mesons is investigated
under the assumption of dominance of the handbag mechanism considering
both quark helicity flip and non-flip. The partonic subprocess, meson
photoproduction off quarks, is analysed with the help of a covariant
decomposition of the subprocess amplitudes which is independent of a 
specific meson generation mechanism. As examples of subprocess
dynamics, however, the twist-2 as well as two-particle twist-3 
contributions are explicitly calculated. Characteristic features of the 
handbag approach are discussed in dependence upon the relative
magnitudes of the invariant functions. Differential cross
sections and spin correlations are predicted to show a characteristic
behaviour which allows to test the underlying assumption of handbag
dominance.
\end{abstract}

\newpage
\section{Introduction}
Recently, the importance of the handbag mechanism in hard exclusive
reactions has been realised.  The handbag mechanism is 
characterised by the fact that only one quark from the incoming and one
from the outgoing nucleon participate in the hard subprocess, while all
other partons are spectators. Factorisation properties have been shown
to hold for Compton scattering  and for meson electroproduction in
the both kinematical regions: the deep virtual region 
characterised by a large virtuality $Q^2$ of the incoming photon and
a small squared invariant momentum transfer $t$ from the incoming to 
the outgoing proton, and the wide-angle region where $-t$ (and $-u$) 
are regarded as the large scale, while $Q^2$ is less than $-t$.
It is to be emphasised that for deep virtual processes, all order
proofs of factorisation exist, while for the wide-angle region
factorisation has only been shown to hold to next-to-leading order in
Compton scattering and to leading order in photo- and
electroproduction of mesons as yet. 

\begin{figure}
\centerline{\includegraphics[width=5.5cm]{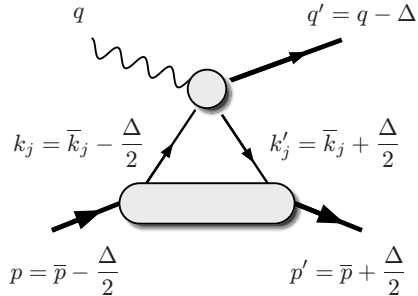}}
\caption{\label{fig:handbag} The handbag diagram for photo- and
electroproduction of mesons. The large blob represents a baryon GPD,
while the small one stands for meson photo- and electroproduction off
partons. The momenta of the various particles are indicated.} 
\end{figure}

As illustrated in Fig.\ \ref{fig:handbag}, the amplitudes for hard
exclusive processes factorise into a parton-level subprocess, e.g.,
meson photo- or electroproduction off quarks, $\gamma^{(*)} q\to M q$, and
generalised parton distributions (GPDs) \ci{mueller} describing the
soft hadron-parton transitions. 
Factorisation is particularly simple in the wide-angle region. 
Instead of convolution as occurring in deep
virtual processes, the wide-angle amplitudes appear as products of
subprocess amplitudes and $t$-dependent form factors which represent
$1/\xb$-moments of GPDs.  
There is another difference between  deep virtual and wide-angle
electroproduction of mesons. The deep virtual process \ci{rad96}   
is dominated by contributions from longitudinally polarised virtual
photons for $Q^2\to \infty$; those from transversally polarised
photons are suppressed by $1/Q^2$. For these amplitudes, factorisation
breaks down \ci{man99}. For wide-angle electroproduction, both
photon polarisations contribute to the same twist order,
there is no break-down of factorisation \ci{huang00}; the limit
$Q^2\to 0$ is therefore unproblematic.

\begin{figure}
\centerline{\includegraphics[width=0.8\textwidth]{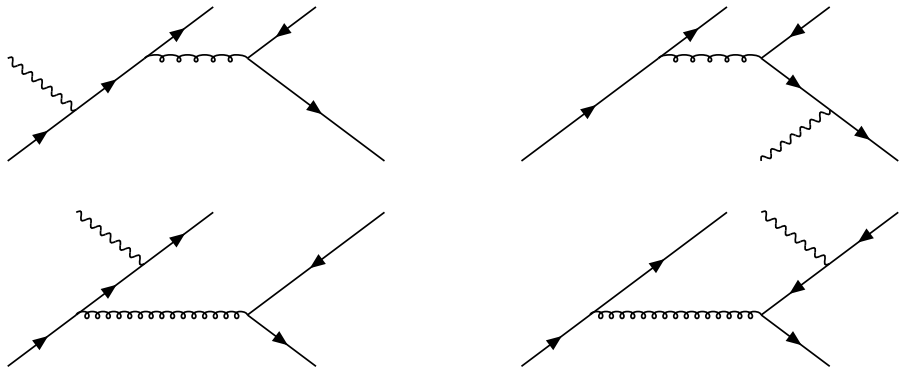}}
\caption{\label{fig:graphs} Lowest order Feynman graphs contributing
to the subprocess $\gamma^{(*)} q\to M q$ to leading twist. The upper
quark and antiquark lines enter the meson's wave function. The
internal curly lines represent hard gluons.}
\end{figure}

Whereas for Compton scattering reasonable agreement with experiment has
been found despite the rather low values of $Q^2$ or $-t,-u$ at which
data is available, in meson electroproduction the magnitude of the
cross sections calculated within the handbag approach
turns out to fail in describing the data~\ci{huang00,man98}. 
The reason for this defect is presumably the one-gluon exchange
mechanism for the generation of the meson (see Fig.\ \ref{fig:graphs})
and not the handbag mechanism itself. Although the one-gluon exchange
mechanism combined with the leading-twist meson distribution amplitude 
dominates for asymptotically large scales, it may fail in
the kinematical situation accessible to current experiments which is
characterised by scales of a few $\gev^2$ only. The lowest-order
Feynman graphs contributing to $\gamma^{(*)} q\to M q$ are the same
as those occurring in the leading-twist calculations of the meson form
factor \ci{rad}. As is well-known (see, for instance \ci{jakob,pang}),
the leading-twist results for the pion form factor are by a factor 3 to 4
below the admittedly poor data available at present \ci{bebek}.
In semi-exclusive pion photoproduction~\cite{Eden:2001ci} which bases 
on the same subprocess $\gamma^{(*)} q\to M q$ too, the normalisation 
also fails in comparison with experiment. 
In the light of these results, a failure by order of magnitude for the 
photo- and electroproduction cross sections is to be anticipated.
Thus, it is probably insufficient to consider 
the one-gluon exchange mechanism for which only zero quark-antiquark 
spatial separations are taken into account and the quark transverse 
momentum is neglected.
It entails a factor $f_M^2/s$ in the photo-
and electroproduction cross section, which is indeed a tiny number at 
large $s$ ($f_M$ is the meson's decay constant). 

How can this situation be remedied ? Suppose handbag factorisation 
continues to hold. Then one may think, for instance, of higher-twist 
or power corrections to the meson generation, resummation of 
perturbative corrections or long-distance meson wave function
effects, which may enhance the leading-twist amplitude
decisively. For instance, by insertion of an infinite number of 
fermionic loops in the hard gluon propagator and interpreting the 
ambiguities in the resummation of these loop effects (known as infrared 
renormalons) as a model of higher-twist contributions an enhancement 
of a factor 3 to 4 is estimated~\cite{bel}. 

In a more pessimistic scenario, one would consider substantial
non-factorising contributions to photo- and electroproduction of
mesons at moderately large values of the hard scale. 
The purpose of this work is to investigate the first scenario, i.e.,
assuming the validity of the handbag mechanism for photoproduction of
pseudoscalar mesons, but allowing for more general subprocess
amplitudes than the one-gluon exchange ones. 
The generalisation of our investigation to electroproduction in the
wide-angle and the deep virtual region is straightforward.

The plan of the paper is the following: In Sect.~\ref{sec:handbag} we 
recapitulate
the handbag mechanism in photoproduction of pseudoscalar mesons ($P$),
extend it in order to include more general mechanisms for the meson 
generation and present some kinematical details. The subprocess 
$\gamma q \to P q$ is discussed in terms of a covariant decomposition
in Sect.~\ref{sec:sub}, and the twist-2 and twist-3 contributions 
to the subprocess are calculated. 
Sect.~\ref{sec:predictions} is devoted to the discussion of characteristic 
predictions 
for meson photoproduction from the handbag approach which may hold 
even if the normalisation of the cross section is not yet understood. 
The paper ends with a summary. 
\section{The handbag mechanism}
\label{sec:handbag}
The handbag mechanism for wide-angle scattering reactions was first
developed for Compton scattering \ci{rad98,DFJK1} and
subsequently applied to photo- and electroproduction of mesons
\ci{huang00}. In this paper we will restrict ourselves to
photoproduction of pseudoscalar mesons, $\gamma B_1\to P B_2$,
where $B_i$ denotes a member of the lowest-lying baryon octet.
For the sake of legibility, we consider here
the case of charged and uncharged pions, and, hence, of nucleons,
leaving the 
generalisations to other pseudoscalar mesons to the end of the paper.
A prerequisite to the application of the handbag
mechanism are Mandelstam variables, $s,\, -t,\, -u$, that are large as
compared to $\Lambda^2$, where $\Lambda$ is a hadronic scale of order
$1 \gev$. It is of advantage to work in a symmetrical frame which is a
c.m.s. rotated in such a way that the momenta of the incoming ($p$) and
outgoing ($p'$) baryons have the same light-cone plus components:
\be
p=\Big[p^+, \frac{m^2-t/4}{2p^+}, -\frac12 \vd\Big]\,, \qquad 
p'=\Big[p^+, \frac{m^2-t/4}{2p^+}, \phantom{-}\frac12\vd\Big]\,,
\ee
where $m$ is the mass of the nucleon, for the definition of other momenta 
see Fig.~\ref{fig:handbag}. The chief advantage of the symmetric 
frame is that the skewness
\be
\xi= \frac{(p-p')^+}{(p+p')^+}\,\
\ee
vanishes.
The baryonic blob in the handbag, see Fig.\ \ref{fig:handbag}, is
viewed as a sum over all possible parton configurations as in deep
inelastic lepton-nucleon scattering (DIS). The crucial assumptions in the
handbag approach are that  of restricted parton virtualities
$k_i^2<\Lambda^2$, and of intrinsic transverse 
parton momenta ${\bf k_{\perp i}}$, defined with respect to their
parent hadron's momentum, which satisfy $k_{\perp i}^2/x_i
<\Lambda^2$, where $x_i$ is the momentum fraction 
that the parton $i$ carries.   
 
One can then show \cite{huang00,DFJK1} that the subprocess Mandelstam 
variables $\sh$ and $\uh$ are the same as the ones for the
full process, photoproduction off baryons, up to corrections of order
$\Lambda^2/t$ ($\th=t$):
\ba
\sh=(k_j+q)^2 \simeq (p+q)^2 =s\,, \quad 
\uh=(k_j-q')^2 \simeq (p-q')^2 =u \,,
\label{eq:hand-ident}
\ea
where $k_j$ denotes the momentum of the active parton, i.e.,\ the one to which 
the photon couples. The active partons are
approximately on-shell, move collinear with their parent hadrons and
carry a momentum fraction close to unity, $x_j, x_j' \simeq 1$.
Thus, like in deep virtual exclusive scattering, the physical
situation is that of a hard parton-level subprocess 
$\gamma q_a\to P q_b$ 
and a soft emission and reabsorption of quarks from the baryon.
The light-cone helicity amplitudes \cite{diehl01} for wide-angle
photoproduction then read 
\ba
\lefteqn{{\cal M}_{0+,\,\mu +}(\gamma B_1\to P B_2)}
\nn \\[0.5em]
 &=& \; \frac{e}{2}
   \sum_{a,b=u,d,s} 
   \left[\, {\cal H}^{P(ab)}_{0+,\,\mu+}(s,t)\,
         \big(R^{\,ab}_{V,B_1\to B_2}(t) + R^{\,ab}_{A,B_1\to B_2}(t)\big)\, 
\right.\nn\\[0.5em]
 && +\left. {\cal H}^{P(ab)}_{0-,\,\mu-}(s,t)\, 
         \big(R^{\,ab}_{V,B_1\to B_2}(t) - R^{\,ab}_{A,B_1\to B_2}(t)\big) 
                                \right] \,,
\label{ampl}
\ea
and
\ba
 \lefteqn{{\cal M}_{0-,\,\mu +}(\gamma B_1\to P B_2)} \nn\\[0.5em]
 &=& \;\frac{e}{2} \frac{\sqrt{-t}}{2m} 
         \sum_{a,b=u,d,s}  
      \left[\, {\cal H}^{P(ab)}_{0 +,\,\mu+}(s,t)\, 
         + \,  {\cal H}^{P(ab)}_{0-,\,\mu-}(s,t)\, \right] 
      \,R^{\,ab}_{T,B_1\to B_2}(t) \,,
\label{eq:lcamp}
\ea
where $\mu$ denotes the helicity of the photon and $e$ is the positron
charge. 
The frame has been chosen such that the momentum transfer $\Delta^\mu$ 
has a positive 1-component and a zero 2-component.
The helicities of the baryons in $ {\cal M}$ and of the quarks 
in the hard scattering amplitude ${\cal H}$ are labelled by their signs 
only for the sake of legibility. The amplitudes for other helicity 
configurations follow from parity invariance:
\be
 {\cal M}^P_{0\nu',\,\mu\nu} = (-1)^{\nu-\nu'}\, 
                           {\cal M}^P_{0-\nu',\,-\mu-\nu}\,.
\label{eq:parity}
\ee
An analogous relation holds for the parton-level amplitudes ${\cal H}$. 

\begin{figure}
\centerline{\includegraphics[width=5.7cm]{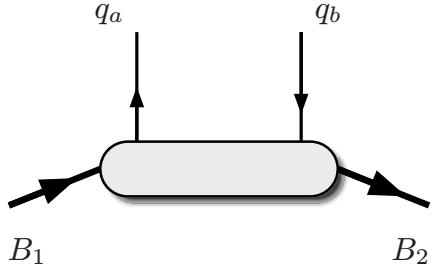}}
\caption{The GPD and the respective form factors
  $R^{\,ab}_{i,B_1\to B_2}$ for $B_1\to B_2$ transitions due to
  emission of a quark with flavour $a$ and reabsorption of a quark with
  flavour $b$.}
\label{fig:blob}
\end{figure}

The hard scattering amplitudes can be computed from the leading-order
Feynman graphs shown in Fig.\ \ref{fig:graphs} and the twist-2 meson
distribution amplitude. 
As a consequence of this leading-twist generation of the meson and the   
masslessness of the quarks there is no quark helicity flip.
A more general mechanism for the generation of the meson, which we 
discuss in the next section, may allow quark helicity flips. 
The form factors $R^{\,ab}_{i,B_1\to B_2}$ represent $1/\xb$-moments of
GPDs at zero skewness, where $\xb=(k_j+k'_j)^+/(p+p')^+$ is the average 
momentum fraction the two quarks carry. The form factors parameterise 
the soft physics that 
controls the emission of a quark with flavour $a$ and the reabsorption of a 
quark with flavour $b$ by a baryon (see Fig.\ \ref{fig:blob}). 
The flavours of the emitted and reabsorbed quarks
fix the quantum numbers of the final baryon uniquely for given initial
baryon. The representation \req{ampl}, \req{eq:lcamp}, which requires 
the dominance of the plus components of the baryon matrix elements, is 
a non-trivial feature given that, in contrast to DIS and deep virtual 
exclusive reactions, not only the plus components of the nucleon
momenta, but also their minus and transverse components are large in this case
\ci{DFJK1}. 

For the proton-proton transitions we use the simplifying
notation ($i=V,A,T$)
\be
    R^{\,aa}_{\,i,p\to p}(t)\equiv R^{\,a}_{\,i}(t)\,.
\ee
These form factors are related to the zero skewness proton 
GPDs \ci{mueller} by
\ba
R^{\,a}_V(t) &=& \int_{-1}^1 \frac{d\xb}{\xb}\, {\rm sign}(\xb)\, 
                                           H^{\,a}(\xb,0;t)\,,\nn\\
R^{\,a}_A(t) &=& \int_{-1}^1 \frac{d\xb}{\xb}\,  
                               \widetilde{H}^{\,a}(\xb,0;t)\,,\nn\\
R^{\,a}_T(t) &=&  \int_{-1}^1 \frac{d\xb}{\xb}\, {\rm sign}(\xb)\,
                                               E^{\,a}(\xb,0;t) \,,
\label{formfactors} 
\ea
where we have used Ji's notation for the GPDs. Note that $\xb$ runs from 
-1 to +1. As usual, a parton with a negative momentum fraction is reinterpreted
as an antiparton with a positive momentum fraction. The GPD 
$\widetilde{E}^a$ and its associated form factor decouples in the
symmetric frame. The form factor $R^{\,a}_T$, which was ignored 
in the previous work on wide-angle photoproduction \ci{huang00}, is 
suppressed by at least $\Lambda/\sqrt{-t}$ as compared to the other 
two form factors. The form factors $R^{\,a}_{\,i}$ for the production 
of pseudoscalar mesons are similar but not identical to those
appearing in wide-angle Compton scattering \ci{DFJK1,HKM}; for instance, 
as a consequence of charge conjugation symmetry, active unpolarised 
quarks and antiquarks contribute with opposite sign in photoproduction of 
pseudoscalar mesons but with the same sign in Compton scattering and 
photoproduction of vector mesons. The GPDs, describing the properties of the 
proton, are the same in both processes, as follows from the 
universality property of the GPDs \ci{mueller}. 

The form factors for other baryon transitions can be related to the
$p\to p$ ones by flavour symmetry. Thus, as a consequence of isospin
invariance, one has ($i=V,A,T$)
\be
R^{\,dd}_{\,i,n\to n}=R^{\,u}_{\,i}\,; \quad R^{\,uu}_{\,i,n\to n}
                                 =R^{\,d}_{\,i}\,; 
       \quad  R^{\,ss}_{\,i,n\to n}=R^{\,s}_{\,i}\,.
\ee
For photoproduction of charged pions, flavour non-diagonal
GPDs occur. These GPDs and, hence, the associated form factors are 
related to the diagonal $p\to p$ ones by \ci{frank99}   
\be
R^{\,ud}_{\,i,p\to n\phantom{+}} \;\equiv\; R_i^{\pi^+} \;=\;
R^{\,du}_{\,i,n\to p\phantom{+}} \;\equiv\; R_i^{\pi^-} \;=\; 
  R^{\,u}_{\,i} - R^{\,d}_{\,i} \,.
\label{eq:isovector}
\ee
As we see, only the
isovector combination contributes to $\pi^\pm$ photoproduction. The
axial form factor $F_A$ \ci{DFJK1} and $R_A$ for ${\pi^\pm}$
photoproduction are both related to the isovector combination of 
$\widetilde{H}$. Owing to the different moments that the form factors
represent, in addition to the valence quarks,
$u$ and $d$ sea quarks contribute to $F_A$, 
whereas they cancel in $R_A$. 

The photoproduction amplitudes for quark helicity flip can be derived in
a way fully analogous to the derivation of the amplitudes \req{ampl}
\ci{huang00}. The result is
\ba
\label{eq:quark-flip-ampl}
{\cal M}^{\,T}_{0+,\,\mu+}(\gamma B_1\to PB_2) &=& 
 {}- e\, \frac{\sqrt{-t}}{2m}\,
             \sum_{a,b=u,d,s}  \left[{\cal H}^{P(ab)}_{0-,\,\mu +}(s,t)- 
                 {\cal H}^{P(ab)}_{0+,\,\mu -}(s,t) \right]\, \nn\\
	  & &\times \,    \Big[S_{S,B_1\to B_2}^{\,ab}(t) + 
               \frac12\, S_{V,B_1\to B_2}^{\,ab}(t)\Big]  
\nn\\
{\cal M}^{\,T}_{0-,\,\mu+}(\gamma B_1\to PB_2)&=& e\, \sum_{a,b=u,d,s} 
                \left[{\cal H}^{P(ab)}_{0-,\,\mu +}(s,t)\,  
                            S_{T,B_1\to B_2}^{\,ab}(t) \right. \nn\\
   && - \left. \frac{t}{4m^2}
       \Big[{\cal H}^{P(ab)}_{0-,\,\mu +}(s,t)
          - {\cal H}^{P(ab)}_{0+,\,\mu -}(s,t)\Big] \, 
       S_{S,B_1\to B_2}^{\,ab}(t)
                          \right] \,.
\ea
These amplitudes are to be added to those given in \req{ampl}. 
Again we denote the form factors for $p\to p$ transitions by
$S_i^{\,a}$. Analogously to the form factors $R^{\,a}_{\,i}$, they
represent $1/\xb$ moments of quark helicity flip distributions
\ba
S_{T}^{\,a} &=&  \int_{-1}^1 \frac{d\xb}{\xb}\,{\rm sign}(\xb)\,
                                                     H^a_T(\xb,0;t)\,, \nn\\
S_{ S}^{\,a} &=& \int_{-1}^1 \frac{d\xb}{\xb}\,{\rm sign}(\xb)\, 
                                          \widetilde{H}^a_T(\xb,0;t)\,,\nn\\
S_{V}^{\,a} &=& \int_{-1}^1 \frac{d\xb}{\xb}\,{\rm sign}(\xb)\, 
                                                    {E}^a_T(\xb,0;t) \,,
\label{eq:trans-formfactors}
\ea
for $i=T,S,V$. The signum functions which control the relative signs
between quark and antiquark contributions, manifest the
charge-conjugation properties of helicity-flip distributions. 
The form factors for $B_1\to B_2$ transitions can be related to the
$p\to p$ ones in the same fashion as for $R_i$, see \req{eq:isovector}. 
The new proton GPDs are defined in \ci{diehl01} (see also \ci{hood98}):
\ba
&&\frac12\, \int\, \frac{dz^-}{2\pi}\, {\rm e}^{i\xb\,\overline{p}^+\,z^-}\; 
               \langle p',\lambda'| \overline{\Psi}_a (-\zb/2)
	       i\sigma^{+i} \Psi_a(\zb/2) |p,\lambda\rangle \nn\\
&&=\frac{1}{\overline{p}^{\,+}}\, \ub(p',\lambda') 
                      \left[\, H^a_T(\xb,\xi;t)\, i\sigma^{+i}
                       + \widetilde{H}^a_T(\xb,\xi;t)\,
         \frac{\overline{p}^{\,+} \Delta^i-\Delta^+
	   \overline{p}^{\,i}}{m^2}\right.\nn\\
&& \quad \left.+ E^a_T(\xb,\xi;t)\, 
         \frac{\gamma^+\Delta^i - \Delta^+\gamma^i}{2m}
               + \widetilde{E}^a_T\,
         \frac{\gamma^+\overline{p}^{\,i}-\overline{p}^{\,+}\gamma^i}{m}
                  \right]\, u(p,\lambda)\,,
\ea
where $i=1,2$ is a transverse index,
\be
\overline{p}=\frac12\, (p+p')\,,
\ee
and $\zb=[0,z^-,\vo]$. At $\xi=0$ the GPD $\widetilde{E}^a_T$ vanishes.
As can be seen from the helicity configurations of the involved quarks
and baryons~\cite{diehl01}, in contrast to $H^a_T$ the GPDs $E^a_T$ 
and $\widetilde{H}^a_T$ involve components of the baryon wave functions
where the parton helicities do not add up to the helicities of the baryons. In
other words, parton configurations with non-zero orbital angular momentum
contribute to it. As inspection of the parton-hadron helicity matrix
elements~\cite{diehl01} reveals, these GPDs as well as the corresponding
form factors $S^a_V$ and $S^a_S$ are expected to be suppressed as compared to 
$S^a_T$ by at least $1/\sqrt{-t}$ and $1/t$, respectively. In this sense the
situation is analogous to the expected suppression of the Pauli form factor 
with respect to the Dirac one, which should be at least 
$\propto 1/\sqrt{-t}$.
On the other hand, there is no argument why $S_T^a$ could be suppressed 
as compared to $R_V^a$. 
Thus, quark helicity flip is suppressed by the subprocess
amplitudes only. It is worth mentioning that all GPDs, those for quark
helicity flip as well as those for non-flip, are real valued, as a 
consequence of time reversal invariance.

\section{The parton-level subprocess $\gamma q_a \to P q_b$}
\label{sec:sub}
\subsection{A covariant decomposition}
The helicity amplitudes for pseudoscalar meson photoproduction off
quarks, $\gamma q_a\to P q_b$, can be decomposed covariantly as
\be
{\cal H}_{0\lambda',\,\mu\lambda}^{P(ab)}(\sh,\th) = \sum_{i=1}^4 
   C^{P(ab)}_{\,i}(\sh,\th)\,\, \ub(k',\lambda')\, Q_i(\mu)\, u(k,\lambda)\,,
\label{eq:dec}
\ee 
where $\sh$, $\uh$, and $\th$ denote the subprocess Mandelstam
variables. 
A suitable set of covariants was given by Chew et al. (CGLN)
\ci{CGLN}:
\ba
Q_1 &=& 2\gamma_5\, \Big[q'\cdot q\; \kb \cdot \veps\, - 
                          q'\cdot\veps\; \kb\cdot q\Big]\,,\nn\\
Q_2 &=& 2\gamma_5\, \Big[\kb\cdot q\; \veps\sla\, - \kb \cdot \veps\; 
                           q\sla\Big]\,,\nn\\
Q_3 &=& \gamma_5\, \Big[q'\cdot q\;\veps\sla\,-  q'\cdot\veps\; 
                           q\sla\Big]\,,\nn\\   
Q_4 &=& \gamma_5\, \veps\sla\, q\sla\,,
\label{eq:cov}
\ea
where $\veps$ is the polarisation vector of the photon. The covariants
$Q_i$ are manifestly gauge invariant and encode the helicity structure 
of the process, whereas the invariant functions $C^{P(ab)}_{\,i}$ depend
on the dynamics. In the symmetric frame, the various momenta and the
photon's polarisation vector have the following representation:
\ba
 & &
\begin{array}{lcl}
\kb = \big[ k^+, k^-, \vo \big] \, , & \quad &
\Delta = \big[ 0, 0, \sqrt{-\th} \,\vea \big]\,, \\[0.2cm] 
q = \big[ k^-, k^+, \frac12 \sqrt{-\th}\, \vea \big]\,, & &
q' = \big[ k^-, k^+, -\frac12 \sqrt{-\th}\, \vea \big]\,, 
\end{array}
\nn \\[0.2cm] & &
\veps(\mu) = \Big[ \frac{\mu}{2} \sqrt{\frac{-\th}{\sh}}\,,
                    - \frac{\mu}{2} \sqrt{\frac{-\th}{\sh}}\,,
          \frac{\mu}{\sqrt{2}}\,\sqrt{\frac{-\uh}{\sh}}\,\vea
	    -\frac{i}{\sqrt{2}}\, \veb \Big]\,,
\label{eq:momenta}
\ea
where the plus and minus components of the average parton momentum 
$\overline{k}=(k_j+k_j')/2$ are given by
\be
k^+= \frac{1}{2\sqrt{2}}\, \big[\sqrt{\sh} + \sqrt{-\uh}\big]\,, \qquad
k^-= \frac{1}{2\sqrt{2}}\, [\sqrt{\sh} - \sqrt{-\uh}]\,.
\label{eq:LCcomp}
\ee

Working out the matrix elements $\ub Q_i u$ in the symmetric frame 
(results are listed in Table~\ref{tab:matrix})
and ignoring quark and meson masses, we can express the parton-level
helicity amplitudes in terms of the invariant functions $C_i$: 
\ba
{\cal H}_{0+,++}^{P(ab)}&=&\phantom{-}\sqrt{-\th/2}\; \sh\, 
                           \Big[C^{P(ab)}_{\,2}+C^{P(ab)}_{\,3}\Big]\,, \nn\\
{\cal H}_{0+,-+}^{P(ab)}&=& -\sqrt{-\th/2}\; \uh\, 
                           \Big[C^{P(ab)}_{\,2} - C^{P(ab)}_{\,3}\Big]\,, \nn\\
{\cal H}_{0-,++}^{P(ab)}&=& -\sqrt{-\uh\sh/2}\; \Big[\th\, C^{P(ab)}_{\,1} -
                                          2\,C^{P(ab)}_{\,4}\Big]\,, \nn\\
{\cal H}_{0-,-+}^{P(ab)} &=& \phantom{-} \sqrt{-\uh\sh/2}\; \th\, 
                                 C^{P(ab)}_{\,1}\,.
\label{eq:sub-ampl}
\ea
Other helicity amplitudes follow from Eq. \req{eq:parity}.
As inspection of \req{eq:sub-ampl} reveals, the invariant functions
$C_2$ and $C_3$ contribute only to the quark helicity conserving
amplitudes, while $C_1$ and $C_4$ generate quark helicity flips. As we
discussed in Sect.~\ref{sec:handbag}, this feature entails different 
proton matrix elements, and hence different form factors, in the full
process $\gamma B_1\to P B_2$. 

\renewcommand{\arraystretch}{2}
\begin{table}
\begin{center}
\begin{tabular}{c||c|c}
      & $\lambda'=\lambda$ & $\lambda'=-\lambda$ \\ \hline\hline  
$Q_1$ & $0$ & $- \sqrt{-\uh \sh/2} \; \th \; \mu$\\
$Q_2$ & $\frac12 \sqrt{-\th/2}\, \Big[\,(1+2\lambda\mu)\,\sh -
    (1-2\lambda\mu)\,\uh\,\Big]$ & $0$ \\
$Q_3$ & $\frac12\sqrt{-\th/2}\,  \Big[\,(1+2\lambda\mu)\,\sh +
    (1-2\lambda\mu)\,\uh\,\Big]$ & $0$ \\
$Q_4$ & $0$ & $ \sqrt{-\uh\sh/2}\, (2\lambda+\mu)$
\end{tabular}
\end{center}
\caption{\label{tab:matrix} Matrix elements of the covariants 
$\ub(k',\lambda') \,Q_i\, u(k,\lambda)$ defined in Eq.~(\ref{eq:cov})
evaluated in the symmetric frame specified by Eqs.~(\ref{eq:momenta}) 
and (\ref{eq:LCcomp}).}

\end{table}
\renewcommand{\arraystretch}{1}

\subsection{Subprocess observables}
\label{sec:sub-observables}
As we will see in the next section, in the context of the handbag mechanism 
observables of the full process are often related to the corresponding ones 
for the subprocess in a simple fashion. It is therefore suitable to present 
some of the subprocess observables in terms of the invariant functions first.

An important observable for testing the handbag mechanism is the
correlation between the helicities of the incoming photon and quark 
\ci{HKM,DFJK2}. This observable is defined by
\be
\hat{A}_{LL}^{P(ab)}=\frac{\sum_{\lambda'}\left\{
                    \Big|{\cal H}^{P(ab)}_{0\lambda',\,++}\Big|^2 
                       - \Big|{\cal H}^{P(ab)}_{0\lambda',\,+-}\Big|^2\right\}}
                    {\sum_{\lambda',\,\lambda}\, 
              \Big|{\cal H}^{P(ab)}_{0\lambda',\,+\lambda}\Big|^2}\,, 
\label{eq:all-def}
\ee
and reads
\ba
\hat{A}_{LL}^{P(ab)}&=& 
              \left\{ (\sh^2-\uh^2)\,\Big[\big|C^{P(ab)}_{\,2}\big|^2 
                         +\big|C^{P(ab)}_{\,3}\big|^2\Big] \right.\nn\\ 
           &&\left. + \, 2\, (\sh^2+\uh^2)\, {\rm Re}\,\big(C^{P(ab)}_{\,2} 
                              C^{P(ab)*}_{\,3}\big)\right. \nn\\
             && \left.+ \, 4\,\frac{\uh\,\sh}{\th}\,
	   \big|C^{P(ab)}_{\,4}\big|^2 -4\, \uh\,\sh\, 
                {\rm Re}\,\big(C^{P(ab)}_{\,1} C^{P(ab)*}_{\,4}\big) 
                 \right\}/\hat{\cal K}^{P(ab)} \,,
\label{eq:all-inv}
\ea
in terms of the invariant functions. Here we have introduced the quantity
\ba
\hat{\cal K}^{P(ab)}&=& (\sh^2+\uh^2)\,\Big[\big|C^{P(ab)}_{\,2}\big|^2 
                   +\big|C^{P(ab)}_{\,3}\big|^2\Big]
    +2\, (\sh^2-\uh^2)\, {\rm Re}\,\big(C^{P(ab)}_{\,2} C^{P(ab)*}_{\,3}\big)
\nn\\
           &+&2\,\th\,\uh\,\sh\, \big|C^{P(ab)}_{\,1}\big|^2 
           + 4\,\frac{\uh\,\sh}{\th}\,
	   \big|C^{P(ab)}_{\,4}\big|^2 - 4\,\uh\,\sh\, {\rm Re}\,
           \big(C^{P(ab)}_{\,1} C^{P(ab)*}_{\,4}\big)\,,
\label{def-K}
\ea
which, up to a factor, represents 
the sum over the squared absolute values of
the subprocess helicity amplitudes.

One may also consider sideways polarisations of the quark where the
$S$ direction is defined by ${\bf S}^{(}{}'{}^) = {\bf N} \times
{\bf L}^{(}{}'{}^)$ for the incoming (outgoing) quark. ${\bf
  L}^{(}{}'{}^)$ is the direction of the incoming (outgoing)
quark's three-momentum and ${\bf N}= {\bf L}\times {\bf L}'$ is the
normal to the scattering plane. The correlation between the helicity
of the incoming photon and the sideways polarisation of the incoming
quark reads \ci{HKM}  
\be
\hat{A}_{LS}^{P(ab)}=2\, \frac{{\rm Re}\; 
           \left\{{\cal H}_{0+,\,++}^{P(ab)}{\cal H}_{0-,\,-+}^{P(ab)\,*}
         - {\cal H}_{0+,\,-+}^{P(ab)}{\cal H}^{P(ab)\,*}_{0-,\,++}\right\}} 
     { \sum_{\lambda',\,\lambda} \Big|{\cal H}_{0\lambda',\,+ 
                                               \lambda}^{P(ab)}\Big|^2 }\,.
\label{def-als}
\ee
Expressing it in terms of the invariant functions, one finds
\ba
\hat{A}_{LS}^{P(ab)} &=& 2\, \sqrt{\frac{\uh\,\sh}{\th}}\,
         {\rm Re}\,\left\{\th\, (\sh-\uh)\, C^{P(ab)}_{\,2} C^{P(ab)*}_{\,1}
                  -\th^{\,2}\, C^{P(ab)}_{\,3} C^{P(ab)*}_{\,1} \right.\nn\\
    && \left.+ \, 2\, \uh\, (C^{P(ab)}_{\,2}-
         C^{P(ab)}_{\,3})C^{P(ab)*}_{\,4}\right\}/ \hat{\cal K}^{P(ab)}\,.
\label{eq:als-inv}
\ea
A non-zero $\hat{A}_{LS}^{P(ab)}$ requires both quark helicity flip
(i.e., non-zero $C_1{}$ and/or $C_4{}$) and non-flip 
(i.e., non-zero $C_2{}$ and/or $C_3{}$) amplitudes.

As we see, the observables in general depend on the invariant functions in a
complicated way. If one of the invariant functions dominates the
others, the expressions (\ref{eq:all-inv}, \ref{def-K}, \ref{eq:als-inv}) 
become simple and clear and exhibit differences characteristic of the 
dominating invariant function. If, for instance, either $C_2$ or $C_3$ 
dominates (or if their phases differ by $90^\circ$ and $C_1$, $C_4$ are
negligible), $\hat{A}_{LL}^{P(ab)}$ reduces to 
\be
\hat{A}_{LL}^{P(ab)} \simeq \frac{\sh^2-\uh^2}{\sh^2+\uh^2}\,. 
\label{eq:all23}
\ee
The result \req{eq:all23} coincides with the helicity correlation in
$\gamma q \to \gamma q$, the subprocess occurring in Compton 
scattering \cite{HKM}. 
In other cases, the helicity correlation differs from \req{eq:all23}
drastically. Thus, for instance, if $C_1{}$ dominates, 
$\hat{A}_{LL}^{P(ab)}$ is zero, while a dominant $C_4{}$ leads to
$\hat{A}_{LL}^{P(ab)}=1$. 
How these subprocess results affect the full process, namely 
meson photoproduction off baryons, remains to be discussed.

\subsection{Contributions from twist-2 and twist-3 meson generation}
\label{sec:twist}
As an example of a dynamical scenario, let us discuss the one-gluon
exchange mechanism. The expressions obtained from the Feynman graphs 
shown in Fig.\ \ref{fig:graphs} are to be convoluted with the meson
distribution amplitude. According to Beneke and Feldmann \ci{ben00},
the light-cone projection operator of an outgoing  
pseudoscalar meson in momentum space,
including twist-3 two-particle contributions, reads 
(conveniently rewritten in the notation of \cite{passek})
\ba
{\cal P}_{\alpha \beta,k l}^{P(ab)} &=& \frac{f_P}{2\sqrt{2 N_C}} \;
{\cal C}_P^{a b} \, \frac{\delta_{k l}}{\sqrt{N_C}}
\left\{\frac{\gamma_5}{\sqrt{2}} \, q\sla{}'
  \phi_P(\tau) \, + \, \mu_P \, \frac{\gamma_5}{\sqrt{2}} 
        \left[ \phi_{Pp}(\tau) 
                                                 \right.\right.\nn\\
    &&\hspace{-0.0cm} \left. \left. 
           - i\sigma_{\mu\nu} \frac{q'{}^\mu k'{}^\nu}{q'\cdot k'} 
              \frac{\phi'_{P\sigma}(\tau)}{6}
                    + i \sigma_{\mu\nu} q'{}^\mu
            \frac{\phi_{P\sigma}(\tau)}{6} \frac{
     \partial}{\partial \ell_{\perp\nu}}\right]\right\}_{\alpha \beta}\,,  
\label{pion-proj}
\ea
where $f_P$ is the meson's decay constant; for instance, 
$f_\pi=132 \mev$; $\alpha$ ($a$, $k$) and $\beta$ ($b$, $l$)
represent Dirac (flavour, colour) labels of the quark and antiquark,
respectively.
The parameter $\mu_P$ in \req{pion-proj} is proportional to the chiral
condensate. For the pion, the familiar result is 
$\mu_{\pi}=m_\pi^2/(m_u+m_d) \simeq 2 \gev$ at a scale of 2 GeV. 
In (26), $\ell_\perp$ denotes the transverse momentum of
the quark entering the meson, defined with respect to the meson's
momentum, $q^\prime$. After performing the derivative the collinear limit,
$\ell_\perp=0$, is taken.
Note that in the massless limit we are working in, the two
vectors $q'$ and $k'/(k'\cdot q')$ are light-like and
their space components have opposite sign%
\footnote{In the frame where $q'=(q'^+,0,\mathbf{0}_{\perp})$,
$q'^{\mu}k'^{\nu}/(k'\cdot q')=q'^{\mu}n^{\nu}/(n\cdot q')$
where $n=(0,1,\mathbf{0}_{\perp})$ and $q'\cdot n=q'^+$.}. 
The projector takes into account the familiar
twist-2 distribution amplitude $\phi_P(\tau)$ 
and the two-particle twist-3 ones $\phi_{Pp}(\tau)$ and $\phi_{P\sigma}(\tau)$,
where $\tau$ is the momentum fraction carried by the quark inside the meson. 
In \req{pion-proj}, $\phi'_{P\sigma}$ denotes the derivative of
$\phi_{P\sigma}$ with respect to $\tau$. A complete projector to
twist-3 accuracy would also include three-particle 
contributions~\ci{braun90}.  
Assuming the three-particle distributions to be strictly zero, the
equations of motion fix the twist-3 distribution 
amplitudes \ci{ben00,braun90} to:
\be 
\phi_{Pp}(\tau) \;=\;1\,, \qquad  
\phi_{P\sigma}(\tau) \;=\; 6\,\tau\,(1-\tau)\,.
\label{da-special}
\ee
Although strictly vanishing three-particle distributions cannot be 
justified \cite{ben00}, the form \req{da-special} gives a hint at the 
magnitude of
the twist-3 two-particle contribution. In \cite{braun90,ball} the next 
Gegenbauer coefficients for $\phi_{Pp}$ and $\phi_{P\sigma}$ have been 
calculated exploiting results for the three-particle twist-3 distribution
amplitudes and equations of motion.
 
The explicit calculation of the leading-order twist-2 contribution to
the process $\gamma q_a\to P q_b$ provides the following results for
the invariant functions:  
\ba
C^{P(ab)}_{\,2}{}\Big|_{twist-2} &=& -2\pi \als(\mu_R^2) f_P\,
  {\cal C}_P^{ab}\,\frac{C_F}{N_c}\,   \frac{e_a\uh+e_b\sh}{\th}\, 
             \frac{\langle 1/(1-\tau) \rangle_P+ 
                        \langle 1/\tau\rangle_P}{\uh\sh}\,,\nn\\
C^{P(ab)}_{\,3}{}\Big|_{twist-2} &=& -2\pi \als(\mu_R^2) f_P
      \, {\cal C}_P^{ab}\, \frac{C_F}{N_c}\, \frac{e_a\uh+e_b\sh}{\th}\,
             \frac{\langle 1/(1-\tau) \rangle_P -  
                        \langle 1/\tau\rangle_P}{\uh\sh}\,,\nn\\
C^{P(ab)}_{\,1}{}\Big|_{twist-2} &=& C^{P(ab)}_{\,4}{}\Big|_{twist-2}=0 \,.
\label{eq:LOamp}
\ea
Here, $\mu_R$ is an
appropriate renormalisation scale and $C_F=(N_c^2-1)/(2N_c)$ is the
usual SU(3) colour factor. 
Note that the positron charge $e$ has been
pulled out of the subprocess amplitudes and is written explicitly in
\req{ampl} and \req{eq:quark-flip-ampl}. 
The flavour weight factors 
${\cal C}_P^{ab}$ comprise the flavour structure of the meson. 
For pions they read
\be
{\cal C}_{\pi^0}^{uu}=-\,{\cal C}_{\pi^0}^{dd}=1/\sqrt{2}\,,\qquad 
{\cal C}_{\pi^+}^{ud}={\cal C}_{\pi^-}^{du}=1\,,
\label{eq:factors}
\ee
all other factors are zero, (e.g., ${\cal C}^{ss}_{\pi^0}=0$),  
since the projection operator \req{pion-proj}
implies the valence quark approximation for the meson.
$\langle 1/\tau\rangle_P$ and $\langle 1/(1-\tau) \rangle_P$ are 
moments of the meson's twist-2 distribution amplitude
\be
\langle 1/\tau \rangle_P = \int^1_0\, d\tau\,
                                     \frac{\phi_P(\tau)}{\tau}\,.
\ee
Owing to the evolution of the distribution amplitudes the moments
depend on the factorisation scale $\mu_F$. The pion 
distribution amplitude is symmetric under the replacement  
$\tau \leftrightarrow 1-{\tau}$; hence $C^\pi_3$ is zero as well. For kaons, 
for instance, this symmetry does not hold. 

To leading-order in $\als$ the twist-3 content of the projector
\req{pion-proj} leads to the invariant functions 
\ba
C^{P(ab)}_{\,1} \Big|_{twist-3} &=&  
- 4 \pi \als(\mu_R^2)\, \frac{C_F}{N_C} \, 
 \frac{f_P \,\mu_P}{\th} \, {\cal C}^{ab}_{P}  \int_0^1 d\tau \nn \\[0.1cm]
            &\times& \,\left\{ \frac{e_a}{\sh^2} 
       \left[ -\frac{\phi_{Pp}}{1-\tau}-
\left( \frac{2}{\tau}+\frac{1}{1-\tau} \right)\, \frac{\phi'_{P\sigma}}{6} 
+ \frac{\phi_{P\sigma}}{3\tau^2} \right] \right. \nn \\[0.1cm] 
&+&  \left. \frac{e_b}{\uh^2} \left[ -\frac{\phi_{Pp}}{\tau}   
+ \left( \frac{2}{1-\tau}+\frac{1}{\tau} \right)\, \frac{\phi'_{P\sigma}}{6} 
+ \frac{\phi_{P\sigma}}{3(1-\tau)^2} \right] \right. \nn \\[0.1cm]
&+&  \left. \frac{e_b\th}{\sh\uh^2}\, \left[ \frac{\phi_{Pp}}{\tau}   
+ \left( \frac{2}{1-\tau}+\frac{1}{\tau} \right)\, \frac{\phi'_{P\sigma}}{6} 
- \frac{(1-\tau-\tau^2)}{\tau^2(1-\tau)^2}\, \frac{\phi_{P\sigma}}{3} 
 \right] \right\} \, , 
\label{eq:tw31}
\ea
and
\ba
C^{P(ab)}_{\,4} \Big|_{twist-3} & =&  - 4 \pi \als(\mu_R^2) \,\frac{C_F}{N_C} 
     \, f_P \, \mu_P \,{\cal C}^{ab}_{P} \int_0^1 d\tau \nn \\[0.1cm] 
&\times& \, \left\{ \frac{e_a}{ \sh^2}  
\left[ -\frac{\phi_{Pp}}{1-\tau} - \, \frac{\phi'_{P\sigma}}{6\tau(1-\tau)}
+ \, \frac{\phi_{P\sigma}}{6\tau^2 (1-\tau)}  \right] \right.
                                          \nn \\[0.1cm] 
&& \left. + \frac{e_b}{\uh^2}\, \left[ -\frac{\phi_{Pp}}{\tau}   
+ \, \frac{\phi'_{P\sigma}}{6\tau(1-\tau)}
+ \, \frac{\phi_{P\sigma}}{6\tau(1-\tau)^2} \right]
              \right. \nn \\[0.1cm] 
&& \left. + \frac{e_b\th}{\sh\uh^2}\, \left[ 
 \frac{\phi'_{P\sigma}}{6\tau(1-\tau)}
- \, \frac{1- 2 \tau}{\tau^2(1-\tau)^2} \frac{\phi_{P\sigma}}{6} 
\right] \right\} \, .
\label{eq:tw34}
\ea
Furthermore,
\be
C^{P(ab)}_{\,2} |_{twist-3}  = 
C^{P(ab)}_{\,3} |_{twist-3}  =  0  \, .
\ee

Inserting the distribution amplitudes (\ref{da-special}) into
Eqs.\ (\ref{eq:tw31}) and (\ref{eq:tw34}), we obtain
\be
C^{P(ab)}_{\,1} |_{twist-3} = C^{P(ab)}_{\,4} |_{twist-3} = 0 \, .
\label{eq:tw314res}
\ee
Although \req{eq:tw314res} is a consequence of the special distribution
 amplitudes (\ref{da-special}) we take it as a hint at the smallness of the
 twist-3 contribution. 
The generalisation to more general distribution amplitudes than 
(\ref{da-special}) is beyond the scope of the present paper 
and is left to a forthcoming publication. 
Such a calculation requires the inclusion of three-particle
twist-3 contributions for consistency. Here we only remark that
the distribution amplitudes proposed in \cite{braun90,ball} lead to finite
results for $C^{P(ab)}_{\,1}$ and $C^{P(ab)}_{\,4}$. 
In particular, as can easily be seen,
factorisation holds for them \footnote{This statement is only
correct if the chiral corrections advocated for in \cite{ball}, are neglected.
These terms break factorisation.}.

Perturbative corrections to the twist-2 and twist-3 contributions will not
change the separation of quark helicity flip and non-flip. Each
additional gluon goes along with an even number of $\gamma$
matrices. Therefore, there is no mixing between $C_{2}$ and $C_3$, on
the one side, and $C_{1}$ and $C_4$ on the other side. 

\section{Predictions from the handbag mechanism for\\
 meson photoproduction}
\label{sec:predictions}
\subsection{General remarks}
The twist-2 contribution \req{eq:LOamp} from the handbag mechanism 
to pion photoproduction has been computed in \ci{huang00}. 
Using the asymptotic distribution
amplitude $\phi_\pi(\tau)=6\tau(1-\tau)$, leading to a value of 3 for 
the $1/\tau$
moment, a cross section has been obtained far below experiment. 
We have also seen that the contribution with the twist-3 meson distribution
amplitude does not solve this problem. 
It is suggestive to assume that a more general mechanism, 
unknown at present, is at work for the generation of the meson. 
It may consist of a resummation of perturbative corrections 
and/or a sum over higher
twists or power corrections. There is, of course, also the possibility
that the handbag contribution does not dominate meson photoproduction
at momentum transfer of the order of $10 \gev^2$.

In this section we discuss predictions for meson
photoproduction from the handbag approach without assuming a specific
mechanism for the generation of the meson. 
Supported by our estimates of the twist-3 contributions, 
we conjecture the suppression of quark 
helicity flip. 
In the handbag mechanism the
baryons emit and reabsorb partons which carry momentum fractions close to
unity (see Sect.~\ref{sec:handbag}). From phenomenological and theoretical 
considerations we expect these
partons to be most likely valence quarks of the baryons; sea quarks are
disfavoured. Hence, in order to simplify our analysis further, we assume the
active partons to be valence quarks of the involved baryons.

Experimental verification of various predictions 
we present below, would be a clear hint at the dominance of the 
handbag mechanism and will shed light on the dynamics of the meson 
generation. Provided the form factors $R_i^P$ are sufficiently 
well-known which will likely be the case soon owing to the running 
JLab measurements on Compton scattering and the universality property 
of the GPDs, it might even be possible to determine the invariant 
functions $C_2^P$ and $C_3^P$ directly from experiment. Despite the still 
not understood normalisation of the cross section, we believe that the 
experimental examination of the handbag predictions is a rewarding goal. 

To start with, we consider the flavour dependence of the
invariant functions $C^{P(ab)}_{\,i}$ for the  production of charged pions. 
In order to understand this dependence, it is instructive to
make the isovector $C_i^{(-)}$ and isoscalar $C_i^{(0)}$ content 
of the invariant functions explicit by writing \ci{CGLN}
\ba
C_{\,i}^{(-)}&=& \frac1{2\sqrt{2}}\, \Big[C_{\,i}^{\pi^+} - 
                                               C_{\,i}^{\pi^-}\Big] \, \nn\\
C_{\,i}^{(\,0\,)}&=& \frac1{2\sqrt{2}}\, \Big[C_{\,i}^{\pi^+} + 
                                                 C_{\,i}^{\pi^-}\Big]\,,
\label{eq:iso-amp}
\ea
where the flavour superscripts $ud$, $du$ are omitted since there is only one
subprocess. $C_{\,2}^{(-)}$ is antisymmetric under the $\sh\leftrightarrow
\uh$ crossing, while $C_{\,2}^{(0)}$ is symmetric \ci{CGLN}. Using
these crossing properties and charge factors corresponding to
isovector and isoscalar combinations of $u$ and $d$ quarks, we write 
\footnote{
Since we are working in the massless limit, $(\sh-\uh)/\th$ is, up to
$\sh\leftrightarrow\uh$ symmetric factors, the only
possibility to construct a dimensionless crossing-odd term.}
\ba
C_{\,2}^{(-)}&=&  (e_u-e_d)\, \frac{\sh-\uh}{\th}\, 
                         \frac{c_2(\sh,\th)}{2\sqrt{2}}\,,\nn\\
C_{\,2}^{(\,0\,)}&=&   (e_u+e_d)\, \frac{c_2(\sh,\th)}{2\sqrt{2}}\,,
\label{eq:isospin}
\ea
where we assume isospin independence of the reduced invariant function 
$c_2$. It is symmetric under the $\sh\leftrightarrow\uh$ crossing.
Hence,
\ba
C_{\,2}^{\pi^+}(\sh,\th) =  \frac{e_u\uh + e_d\sh} 
                                       {\sh+\uh}\, c_2(\sh,\th)\,,\nn\\
C_{\,2}^{\pi^-}(\sh,\th) =  \frac{e_d \uh + e_u\sh} 
                                             {\sh+\uh}\, c_2(\sh,\th)\,,
\label{eq:c2pi}
\ea
in conformity with the twist-2 contribution \req{eq:LOamp}. 
The associated form factors $R^{\,ud(du)}_{\,i,p(n)\to n(p)}=R^{\pi^\pm}$
are given in \req{eq:isovector}. 

The invariant function $C_{\,3}^\pi$ exhibits a crossing behaviour
opposite to that of $C_{\,2}^\pi$ \ci{CGLN}. Therefore, we assume in
analogy to \req{eq:isospin}
\ba
C_{\,3}^{(-)}&=&  (e_u-e_d)\, \frac{c_3(\sh,\th)}{2\sqrt{2}}\,,\nn\\ 
C_{\,3}^{(\,0\,)}&=&  (e_u+e_d)\,\frac{\sh-\uh}{\th}\, 
             \frac{c_3(\sh,\th)}{2\sqrt{2}}\,,
\label{eq:isospin3}
\ea
which leads to
\ba
C_{\,3}^{\pi^+}(\sh,\th) &=&  \frac{e_u\uh - e_d\sh} 
                                       {\sh+\uh}\, c_3(\sh,\th)\,,\nn\\
C_{\,3}^{\pi^-}(\sh,\th) &=&   \frac{e_d\uh - e_u \sh} 
                                             {\sh+\uh}\, c_3(\sh,\th)\,.
\label{eq:c3pi}
\ea
The reduced invariant function $c_3$ is crossing symmetric as
$c_2$. 

For $\pi^0$ photoproduction we have to consider two subprocesses
$\gamma u\to\pi^0 u$ and $\gamma d\to\pi^0 d$.
According to~\cite{CGLN} their isospin decomposition involves the isoscalar
part $C_i^{(0)}$, which also appears in the charged pion case, and a new
isovector component $C_i^{(+)}$ which has the same crossing properties as
$C_i^{(0)}$. In analogy to (\ref{eq:isospin}), (\ref{eq:isospin3}) and
assuming again isospin independence of the reduced invariant functions 
we write
\ba
C_{\,2}^{(+)} &=& (e_u-e_d)\, \frac{c_2(\sh,\th)}{2\sqrt{2}} \,,\nn\\ 
C_{\,3}^{(+)} &=& (e_u-e_d)\,\frac{\sh-\uh}{\th}\, 
                   \frac{c_3(\sh,\th)}{2\sqrt{2}}\,. 
\ea 
Sum and difference of $C_i^{(0)}$ and $C_i^{(+)}$ provide the subprocess
amplitudes for the $\pi^0$ photoproduction ($a=u,d$)
\ba
C_{\,2}^{\pi^0(aa)}(\sh,\th) &=& e_a\, {\cal C}_{\pi^0}^{aa}
                                 \,c_2(\sh,\th)\,,\nn\\
C_{\,3}^{\pi^0(aa)}(\sh,\th) &=& e_a\, {\cal C}_{\pi^0}^{aa}
                                 \frac{\sh-\uh}{\th}\,c_3(\sh,\th)\,,
\ea
where the ${\cal C}_{\pi^0}^{aa}$ are the flavour weight factors defined 
in~(\ref{eq:factors}). 
The leading-twist result (\ref{eq:LOamp}) is again in agreement with these
isospin and crossing considerations. Note that in contrast to the case of
charged pions, the flavour dependence here appears as a constant factor. It is
therefore convenient, following previous work~\cite{huang00,man98,DFJK1}, 
to pull out the flavour dependence $e_a\, {\cal C}_{\pi^0}^{aa}$ from the
subprocess amplitudes and to absorb them into the form factors which leads to
\be
R_{\,i}^{\,\pi^0} = e_u\, {\cal C}_{\pi^0}^{uu}\, R^u_{\,i} + 
                    e_d\, {\cal C}_{\pi^0}^{dd}\, R^d_{\,i}  
                  =\frac{1}{\sqrt{2}} ( e_uR_i^u - e_d R_i^d)
\,.
\label{eq:pi0ff}
\ee
The remaining subprocess amplitudes for which the flavour labels are to be
dropped now, read
\be
C_{\,2}^{\pi^0}= c_2(\sh,\th)\,, \qquad C_{\,3}^{\pi^0}=
\frac{\uh-\sh}{\sh+\uh} c_3(\sh,\th)\, .
\label{eq:pi0}
\ee

For subsequent numerical studies, a model for the form factors
or for the underlying GPDs is required.
Following \ci{huang00,DFJK1,HKM}, we use a model which is
motivated by overlaps of light-cone wave functions \ci{rad98,DFJK1,DFJK3}
and is designed for large $-t$ and zero skewness:
\ba
H^a(\xb,0;t) &=& \exp{\left[a^2_N t
        \frac{1-\xb}{2\xb}\right]}\, \big[q_a(\xb)-\bar{q}_a(\xb)\big]\,,\nn\\ 
\widetilde{H}^a(\xb,0;t) &=& \exp{\left[a^2_N t
        \frac{1-\xb}{2\xb}\right]}\, 
           \big[\Delta q_a(\xb) -\Delta \bar{q}_a(\xb)\big]\,,
\label{gpd}
\ea
where $q(\xb)$ and $\Delta q(\xb)$ are the usual unpolarised and polarised
parton distributions in the proton \footnote{ In the full process for
  which the GPDs are needed, $t$ is identified with $\th$.}$^,$\footnote{
In principle, the helicity-flip GPD $H_T^a$ can be modelled analogously
to \req{gpd}. In this case, the corresponding PDFs are the 
transversity distributions $\delta^a(\xb)$.}.
Note that here we display
quarks and antiquarks explicitly; therefore, the moments of the GPDs
are obtained from them by integrating only from $\xb=0$ to 1. Forced
by the Gaussian in \req{gpd}, only large $\xb$ contribute for large
$-t$. For the transverse size parameter of the proton $a_N$, we take 
a value of $0.8\gev$ \ci{DFJK1}. Since the phenomenological parton 
distributions \ci{grv} suffer from large uncertainties at large $\xb$, 
which is an important region for hard wide-angle scattering, the model 
is to be improved at large $\xb$, see \ci{DFJK1,kroll03}. 

The form factor $R_T$ is not modelled explicitly but rather fixed
relative to $R_V$. According to a recent measurement performed 
at JLab \ci{gayou}, its electromagnetic analogue, the Pauli form factor 
of the proton $F_2$, seems to be suppressed by $\Lambda/\sqrt{-t}$ 
as compared to the Dirac form factor $F_1$. Since, on the one hand, 
$F_1$ and $R_V$ are related to the GPD $H$, while, on the other hand, 
$F_2$ and $R_T$ are associated with $E$, it is plausible to expect a 
similar suppression of $R_T$, as seems to occur for $F_2$. Hence, as
has been discussed in \ci{HKM}, the ratio 
\be
\kpt = \frac{\sqrt{-t}}{2m}\; \frac{R^{\,P}_T}{R^{\,P}_V}
\ee
may be expected to be a constant at large $-t$ and $R_T$ and, hence,
the proton helicity flip amplitude \req{eq:lcamp}, has therefore
to be taken into account. However, a suppression of $R_T$ 
like $\Lambda^2/t$ cannot be excluded as yet \ci{arr},
although recent estimates \cite{mel} of two-photon exchange contributions 
seem to favour the JLab results \cite{gayou}. In the former case,
one may ignore $R_T$ and proton helicity flip for not too small
values of $-t$. 
\subsection{Ratio of $\pi^+$ and $\pi^-$ photoproduction cross
  sections}
The derivation of the photoproduction amplitudes within the handbag
approach naturally requires the use of the light-cone helicity
basis. However, for comparison with experimental and other theoretical
results, the use of the ordinary photon-nucleon c.m.s. helicity basis
is more convenient. The standard helicity amplitudes 
$\Phi_{0\nu',\,\mu\nu}$, defined in a c.m.s. in which the photon and 
the incoming proton move along the 3-direction,
are obtained from the light-cone helicity
amplitudes \req{ampl}, defined in the symmetric frame, by the
following transform \ci{HKM,diehl01}:
\be
\Phi^P_{0\nu',\,\mu\nu} = {\cal M}^P_{0\nu',\,\mu\nu} +\beta/2\Big[
        (-1)^{1/2-\nu'}\, {\cal M}^P_{0-\nu',\,\mu\nu} 
       + (-1)^{1/2+\nu}\, {\cal M}^P_{0\nu',\,\mu-\nu}\Big]\,,
\ee
where
\be
\beta= \frac{2m}{\sqrt{s}}\, \frac{\sqrt{-t}}{\sqrt{s}+\sqrt{-u}}\,.
\ee
Since we are now discussing the full process within the handbag
approach, we can make use of Eq.\ \req{eq:hand-ident}. As pointed out 
in Ref.\ \ci{DFHK}, besides this there are ambiguities in relating
the massless kinematics used in the handbag approach with the
experimental one for which, in particular at energies available at
JLab, the proton mass cannot be ignored. For the numerical results
presented here we use the identification 
of our Mandelstam variables with the
experimental ones ($s_{\rm exp}$, $u_{\rm exp}$,  $t_{\rm exp}$) in
which the nucleon mass is taken into account properly 
(scenario 2 in the terminology of \ci{DFHK}) 
\be
s= s_{\rm exp} - m^2\,, \qquad t=t_{\rm exp}\,, \qquad u=u_{\rm exp} -
m^2\,.
\label{eq:kin-id}
\ee

The unpolarised differential cross section for the production of a
pseudoscalar meson is given by
\be
\frac{d\sigma^P}{dt} = \frac{1}{32\pi s^2} 
            \sum_{\nu',\mu} \Big|\Phi^P_{0\nu',\,\mu +}\Big|^2\,,
\label{eq:cross}
\ee
where the sum over the squared helicity amplitudes can be expressed as
\be
\sum_{\nu',\mu} \Big|\Phi^P_{0\nu',\,\mu +}\Big|^2 =
             -\Big(\frac{e}{2}\Big)^2\, t\,(s-u)^2 {\cal K}^P\,.
\ee
For photoproduction of $\pi$ mesons, the function
${\cal K}^P$ is defined by
\ba
{\cal K}^{P}&=&
      \left[\big(R_V^{\,P})^2\,\big[1+(\kpt)^2\big] 
            + \frac{t^2}{(s-u)^2}\, (R_A^{\,P})^2\right]\, 
                                \big|C_{\,2}^{P}\big|^2\nn\\
     &&+ \, \left[ \frac{t^2}{(s-u)^2}\,\big(R_V^{\,P})^2\, 
      \big[1+(\kpt)^2\big] + (R_A^{\,P})^2\right]\, 
                                \big|C_{\,3}^P\big|^2\nn\\  
     &&- \, \frac{t}{s-u}\, \left[\big(R_V^{\,P})^2\,\big[1+(\kpt)^2\big] 
            + (R_A^{\,P})^2\right]\, 
              2\, {\rm Re}\,(C^{P*}_{\,2}  C_{\,3}^{P})\,,
\label{eq:K}
\ea 
if $ C_{\,1}^{P}$ and  $ C_{\,4}^{P}$ are neglected.

If $C_{\,2}^P$ dominates, we obtain from \req{eq:c2pi} and \req{eq:K} for 
the ratio of $\pi^{\pm}$ cross sections \footnote{The invariant
functions  $C_{\,1}^{(0,-)}$ and  $C_{\,4}^{(0,-)}$ have the same
crossing properties as $C_{\,2}^{(0,-)}$ and would therefore lead to
the same ratio of $\pi^-$ and $\pi^+$ cross sections.}
\be 
\frac{d\sigma(\gamma n\rightarrow\pi^-p)}{d\sigma(\gamma p\rightarrow\pi^+ n)}
 = \left(\frac{e_u s + e_d u}{e_u u+e_d s}\right)^2\,.
\label{ratio}
\ee
\begin{figure}
\centerline{\includegraphics[width=0.55\textwidth]{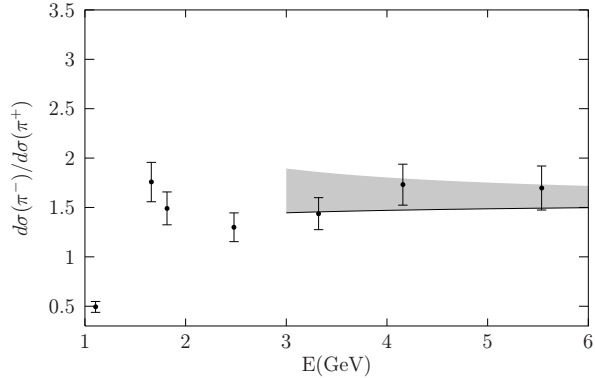}}
\caption{\label{fig:pmratio} The ratio of the $\gamma n\to \pi^- p$ and
$\gamma p \to \pi^+n$ cross sections versus photon beam energy $E$, at a
c.m.s. scattering angle of $90^\circ$. Data are taken from \ci{zhu02}. 
The solid line is the handbag prediction with the identification 
\req{eq:kin-id}. The uncertainties due to target mass corrections
\ci{DFHK} are indicated by the shaded band.}
\end{figure}

As Fig.\ \ref{fig:pmratio} reveals, the prediction \req{ratio} is in
surprisingly good agreement with recent ex\-pe\-ri\-ment\-al results 
from JLab \ci{zhu02} given the small photon beam energies 
($s_{\rm exp}=2mE+m^2$). For comparison, we also quote the result for 
the ratio 
of cross section which follows from the dominance of $C_{\,3}^P$.
In this case, we find from \req{eq:c3pi} and \req{eq:K}
\be
\frac{d\sigma(\gamma n\rightarrow\pi^- p)}{d\sigma(\gamma p\rightarrow\pi^+ n)}
 = \left(\frac{e_u s - e_d u}{e_u u - e_d s}\right)^2\,,
\ee
which tends to infinity at large $s$ and a scattering angle of $90^\circ$
and is clearly at variance with experiment \ci{zhu02}. For the special
case of equal reduced invariant functions, $c_2=c_3$, the ratio of
cross sections is unity. Thus, there is a strong indication from
experiment that the handbag mechanism is at work in these processes 
with $|C_{\,2}^P|\gg |C_{\,3}^P|$ 
provided our assumption of negligible quark helicity flip 
contributions holds.
Structurally, the result (\ref{ratio}) coincides with the
leading-twist prediction
with, however, a more general function $C_{\,2}^P$.

\subsection{Spin correlations}
\label{sec:corr}
Spin correlations provide further severe tests of the handbag mechanism.
They are given by ratios of partial cross sections within one and the same
process and are, therefore, independent of flavour symmetry
breaking effects. If one of the invariant functions
dominates, for which the ratio of $\pi^\pm$ cross sections provides
some evidence, it cancels and an absolute 
prediction for spin correlations is obtained.
  
\begin{figure}
\centerline{\includegraphics[width=0.55\textwidth]{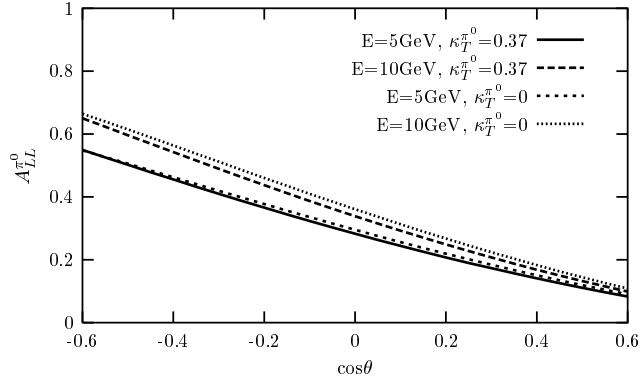}}
\caption{\label{fig:all} Handbag predictions for the helicity
correlation \protect\req{eq:all23f} 
in $\pi^0$ photoproduction 
vs. $\cos\theta$ at two different beam energies and two values of the
ratio $\kappa_{\scriptscriptstyle T}^{\scriptscriptstyle \pi^0}$ 
assuming the dominance of the
invariant function $C_{\,2}^{\pi^0}$. Target mass corrections are
not shown.} 
\end{figure}

\begin{figure}
\centerline{\includegraphics[width=0.55\textwidth]{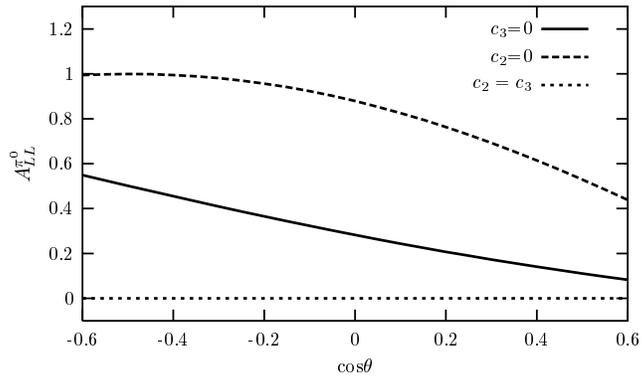}}
\caption{\label{fig:all-2} Handbag predictions for the helicity
correlation $A_{LL}$ \protect\req{eq:all23f}
in $\pi^0$ photoproduction 
vs. $\cos\theta$ at a beam energy of 5 GeV and with
$\kappa_{\scriptscriptstyle T}^{\scriptscriptstyle \pi^0}=0.37$.
For the invariant functions \req{eq:pi0}, the three cases  
$c_3=0$, $c_2=0$, and $c_3=c_2$ are investigated.} 
\end{figure}

The correlation of the incoming photon and nucleon helicities,
$A_{LL}$, or the helicity transfer from the incoming photon to
the outgoing nucleon $K_{LL}$, is defined as in \req{eq:all-def} but with the
subprocess amplitudes, ${\cal H}$, replaced by the corresponding
amplitudes, $\Phi$, of the full process. In terms of form factors and 
invariant functions,
the helicity correlation for pion photoproduction reads 
\ba
A^{P}_{LL} &=& K^{P}_{LL}= \frac{-2t}{s-u}\,
     \frac{R^{\,P}_A \, R^{\,P}_V}{{\cal K}^P}\, (1+\beta \, \kpt)\,
\nn\\ 
           &\times&     \left[\big|C_{\,2}^P\big|^2 +
      \big|C_{\,3}^P\big|^2 +2\frac{s^2+u^2}{s^2-u^2}  
             {\rm Re}\big(C_{\,2}^{P*}\, C_{\,3}^{P}\big)\right]\,,
\label{eq:all23f}
\ea
if $C_1^P$ and $C_4^P$ are neglected.
The expression strongly simplifies if one of the invariant
func\-tions domi\-nates and if some reasonable kinematical approximations
as well as the fact that $R_A^2\leq R_V^2$, are used. The latter
property follows from \req{gpd} directly. Thus, if 
$C_{\,2}^P \gg C_{\,3}^P$, one finds 
\be
A^{P}_{LL}  \simeq \frac{s^2-u^2}{s^2+u^2}\, 
             \frac{R^{\,P}_A}{R^{\,P}_V}\, (1+\beta \, \kpt)\,,
\label{eq:all2}
\ee
while for  $C_{\,3}^P \gg C_{\,2}^P$, the roles of $R^{\,P}_V$ and
$R^{\,P}_A$ are roughly interchanged.
Equation \req{eq:all2} is the same expression as one finds in
wide-angle Compton scattering \ci{HKM}. Provided one of the invariant
functions dominates, the helicity correlation should exhibit a
scattering angle dependence as that of the corresponding observable
for the partonic subprocess \req{eq:all23}; only its magnitude is
diluted by a ratio of slightly process-dependent form factors, see
Fig.\ \ref{fig:all}. If, on the other hand, the reduced invariant 
functions \req{eq:pi0} have roughly the same value, $c_2 \simeq c_3$, 
one obtains $A_{LL}^{\pi^0} \simeq 0$, while the helicity correlations 
for charged pions are non-zero but small in magnitude.
In Fig.\ \ref{fig:all-2} we display results for the helicity
correlation in $\pi^0$ photoproduction obtained for three different 
assumptions on the reduced invariant functions $c_2$ and $c_3$ in order to 
demonstrate the sensitivity of this observable to the underlying 
meson generation mechanism.

\begin{figure}[t]
\centerline{\includegraphics[width=0.55\textwidth]{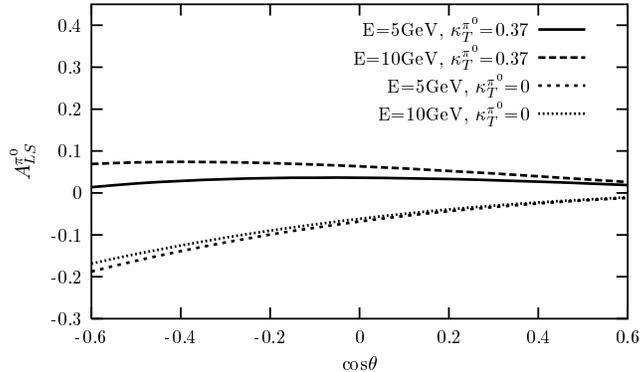}}
\caption{\label{fig:als} 
The helicity correlation $A_{LS}$ \protect\req{eq:als}
for $\pi^0$
photoproduction. For the legend, see Fig.~\ref{fig:all}.} 
\end{figure}

The correlation between the helicity of the incoming photon and the
sideways polarisation of either the incoming ($A_{LS}$) or outgoing ($K_{LS}$)
nucleon is defined as in \req{def-als} but, as for $A_{LL}$, with the
${\cal H}$ being replaced by the amplitudes $\Phi$.
This definition leads to ($C_1^P$, $C_4^P$ neglected) 
\ba
A_{LS}^{P}&=& - K_{LS}^{P} = \frac{2t}{s-u}\, \frac{R_V^{\,P}
                    R_A^{\,P}}{{\cal K}^P}\, (\beta - \kpt)\nn\\
             &&\left[\Big|C_{\,2}^{P}\Big|^2 + \Big|C_{\,3}^P\Big|^2
                       + 2\frac{s^2+u^2}{s^2-u^2} {\rm Re}(C_{\,2}^P
		       C_{\,3}{}^{P*}) \right] \,. 
\label{eq:als}
\ea
Predictions for $A_{LS}$ are shown in Fig.\ \ref{fig:als}.
If $C_{\,2}^P \gg C_{\,3}^P$, for instance, \req{eq:als}
simplifies to
\be
A_{LS}^{P} \simeq \frac{s^2-u^2}{s^2+u^2} \,
            \frac{R_A^{\,P}}{R_V^{\,P}}\, 
            \big(\kpt-\beta\big) \,,
\label{eq:als23}
\ee
and a similar expression in the case of a dominant $C^{P}_{\,3}$.
In contrast to $A_{LL}^P$, which
is a rather robust prediction of the handbag approach, $A_{LS}^P$
depends on the difference of two correction terms: the kinematical factor
$\beta$ which controls the transform of the light-cone helicity amplitudes
to the standard c.m.s.\ ones and the badly known tensor form
factor. Hence, one has to be aware of possible large corrections to
$A_{LS}^P$. We also repeat that the subprocess $\hat{A}_{LS}^{P(ab)}$ 
is zero if quark helicity flip is neglected, see \req{eq:als-inv},
which again signals that $A_{LS}^P$ is dominated by soft physics.

A particular noteworthy result is the ratio of the two correlation
parameters 
\be
\frac{A^P_{LS}}{A^P_{LL}}= -\frac{K^P_{LS}}{K^P_{LL}}= 
       \frac{\kpt  -\beta}{1 + \beta \kpt}\,
\label{eq:lsllratio}
\ee
which holds for any values of $C_{\,2}^P$ and $C_{\,3}^P$ as can
easily be seen from \req{eq:all23f} and \req{eq:als}.
This result is the same as in Compton scattering \ci{HKM}.  
Many corrections, as for instance, those due to the proton mass cancel 
in \req{eq:lsllratio} to a large extent
\ci{DFJK3}. 

Before closing this section a remark regarding the quark helicity-flip
contribution is in order. Its dominance leads to very different
results as we have already noted in Sect.~\ref{sec:sub-observables}. 
Thus, for instance, if we
take only the leading form factor, $S_T$, into account 
(see \req{eq:trans-formfactors}),
the dominance of $C_1^P$ leads to $A_{LL}^P=0$ while a dominant
$C_4^P$ provides  $A_{LL}^P=1$.

At JLAB the observables $K_{LL}$ and $K_{LS}$ have recently been measured
for $\pi^0$ photoproduction~\cite{Wijesooriya:2002uc} in a kinematical 
range however in which the handbag approach cannot be applied. Measurements 
of these observables at higher energies are needed.

\subsection{Generalisation to other pseudoscalar mesons}
\label{sec:gen}
Up to this point we have concentrated on photoproduction of pions.
However, nearly all that we have said holds for other pseudoscalar mesons
as well and is, in fact, written in a form that allows a straightforward
generalisation. Thus, the handbag amplitudes \req{ampl}, \req{eq:lcamp}, 
\req{eq:quark-flip-ampl} (with the nucleon mass replaced by an appropriate 
one) hold directly as well as the
covariant decomposition \req{eq:dec}, \req{eq:sub-ampl}, the
subprocess observables and the twist-2 and twist-3 results. One only
has to insert the relevant form factors and flavour factors 
${\cal C}_P^{ab}$. For the process 
$\gamma p\to K^{+(0)} \Sigma^{0(+)}$, one has for instance 
\be
R^{\,ds}_{\,i,\,p\to \Sigma^+}\simeq \sqrt{2} R^{\,us}_{\,i,\,p\to \Sigma^0}
\simeq 
    - R^{\,d}_{\,i} + R^{\,s}_{\,i}\,, 
\label{eq:RKaon}
\ee
and ${\cal C}_{K^+}^{us}=1$, ${\cal C}_{K^0}^{ds}=1$.
All other flavour factors are zero. 
The form factors \req{eq:RKaon} as well as
those for other processes can easily be obtained 
from the information given in \ci{frank99}. 

The results for the spin correlations discussed in 
Sect.~\ref{sec:corr}, hold for other pseudoscalar
mesons as well. In particular, if $C_{\,2}^P$ dominates in a given process,
then \req{eq:all2} and \req{eq:als23} are approximately valid. 
One has to be aware that the mass of the nucleon in the definition of
the quantity $\beta$ (in \req{eq:lcamp}, \req{eq:quark-flip-ampl}, too) 
has to be replaced appropriately. Also, the ratio of 
$A_{LS}$ and $A_{LL}$ \req{eq:lsllratio}, holds. 
Ratios of cross sections, analogue to \req{ratio}, can also be discussed. 
In this context the situation for kaons is however more intricate. 
Although the reaction
$\gamma N\to\Sigma K$ has an isospin decomposition of the same form as in the
pion photoproduction~\cite{Donnachie}, the invariant functions do not have
definite behaviour under $\sh\leftrightarrow\uh$ crossing. But if we simply 
generalise the leading-twist result (\ref{eq:LOamp}) we obtain the
equivalent of (\ref{eq:c2pi}). From this we find
\be 
\frac{d\sigma(\gamma p\to K^0 \Sigma^+)}{d\sigma(\gamma p\to K^+ \Sigma^0)}
 = 2 \left(\frac{e_d u + e_s s}{e_u u + e_s s}\right)^2\,.
\ee
As for the case of pions, see \req{ratio}, the form factors
\req{eq:RKaon} cancel here.

When comparing photoproduction cross sections for pseudoscalar mesons
not belonging to the same isospin multiplet, flavour symmetry breaking
is to be considered. 
The decay constants play an important role in understanding many
properties of flavour symmetry breaking in the pseudoscalar meson sector
\ci{FKS1}. One may therefore expect that the decay constants also
account for the main symmetry breaking effects in meson photoproduction.
If so, the ratio of the $\eta$ and $\pi^0$ cross sections should, for
instance, approximately be proportional to 
$(f_\eta^{\rm eff}/f_\pi)^2\simeq 2$, 
where the effective $\eta$ decay constant may be evaluated employing 
the quark-flavour mixing scheme advocated for in \ci{FKS1}. In addition, 
there is a minor effect due to the different form factors in the cross 
section ratio. For other cross section ratios, one has to be aware of
additional flavour symmetry breaking effects or, in the case of the
$\eta^\prime$, its glue content may be of importance. 
As an example how to take into account
the flavour mixing together with $gg$ components of the wave function we
present the derivation of the leading-twist result for $\eta$ and $\eta^\prime$
photoproduction in the Appendix.
 
Finally, an interesting peculiarity is to be noted for the leading-twist
calculation in the case of kaons. The kaon distribution amplitude does
not exhibit the symmetry under the replacement $\tau \leftrightarrow
1-\tau$ and therefore has odd and even terms in the Gegenbauer
expansion \ci{CZ} in contrast to the case of the pion. As we see from
\req{eq:LOamp}, $C^K_{3}$ is related to the odd coefficients in the 
Gegenbauer expansion and is therefore not necessarily zero to
leading-twist accuracy. According to the estimates of the lowest 
Gegenbauer coefficients (see, for instance, \ci{bolz}), $C^K_{3}$ seems 
to amount to only  about $10\%$ of  $C^K_{2}$ to twist-2
accuracy. 

\section{Summary}
The handbag mechanism for wide-angle photoproduction of pseudoscalar
mesons has been investigated. In contrast to the analysis performed in
\ci{huang00} where the leading-twist generation of the meson has been
assumed, the partonic subprocess, $\gamma q\to P q$, has here been
treated by means of the CGLN covariant decomposition \ci{CGLN}. The associated
four invariant functions encode the dynamics of the subprocess and in
particular that of the meson generation. This way we can consider
quark helicity flip and non-flip contributions. While in the latter
case the treatment of the handbag is analogue to that occurring in
wide-angle Compton scattering \ci{huang00,DFJK1}, the quark
helicity-flip contribution necessitates the introduction of
helicity-flip GPDs and the associated form factors about which not
much is known at present.

Depending on the relative magnitudes of the invariant functions, the
handbag approach leads to the characteristic predictions for ratios of
cross sections and spin correlations. Provided quark helicity flip can
be neglected, the data \ci{zhu02} on the ratio of the $\pi^-$ and
$\pi^+$ cross sections give a strong indication at the dominance of
the invariant function $C_{\,2}$. It would be interesting to see
whether the data on the helicity correlation support this finding. If so,
we would be tempted to conclude that the handbag mechanism is at work in
meson photoproduction with invariant function for the subprocess which
respects the relative order of magnitude, as predicted by a calculation
of the twist-2 and twist-3 contributions, although their absolute size
must be larger. It remains to be seen then whether resummation of
perturbative corrections or higher twist effects may lead to such
a large invariant function  $C_{\,2}$.   

\section*{Acknowledgements}
  The authors are grateful to Andrei Belitsky, Vladimir Braun, 
Markus Diehl, Thorsten Feldmann, Bla\v{z}enka Meli\'{c}, and
Wolfgang Schweiger for valuable comments.
This work was partially supported by the Ministry of Science 
and Technology of Republic of Croatia under Contract No. 0098002
and by the ESOP network under Contract No. HPRN-CT-2000-00130.
\begin{appendix}

\section{The leading-twist result for $\eta$ and $\eta^\prime$ photoproduction}
\label{sec:eta}
{}From the discussions in the preceding sections it has become clear that
the twist-2 contribution does not dominate wide-angle photoproduction
of pseudoscalar mesons in the kinematical range currently
accessible. Despite of this we will provide the leading-twist result
for $\eta$ and $\eta^\prime$ photoproduction for the sake of completeness
and for possible future use.  

For the $\eta$ and $\eta^\prime$ mesons, mixing is to be taken into
account. It is advantageous for a perturbative calculation to choose 
SU(3)$_F$ singlet and octet flavour combinations of quark-antiquark
states as the valence Fock states of $\eta$ and $\eta^\prime$ 
mesons \ci{passek}. The photoproduction amplitudes are then calculated 
for these combinations separately with corresponding distribution
amplitudes and decay constants $f_\eta^i$, $f_{\eta^\prime}^i$,
$i=1,8$. The numerical values of these decay constants have been 
evaluated in Ref.\ \ci{FKS1}:
\ba
f_\eta^8 &=& \phantom{-}1.17 f_\pi\,, \qquad   f_\eta^1= 0.19 f_\pi\,, \nn\\
f_{\eta^\prime}^8 &=& -0.46 f_\pi\,,  \qquad  f_{\eta^\prime}^1= 1.15 f_\pi\,.
\ea
The photoproduction amplitudes for $\eta$ or $\eta^\prime$
production off protons are the sum of 
the corresponding singlet and octet amplitudes. 
In addition to the $q\bar{q}$ combinations there is also a
two-gluon state which also possesses flavour-singlet quantum numbers and
contributes to leading twist. For electroproduction, the corresponding
amplitude for longitudinally polarised virtual photons has been
calculated in \ci{passek}. Owing to the smallness of  $f_\eta^1$ 
the two-gluon contribution to $\eta$ photoproduction is unimportant, in
contrast to $\eta^\prime$ production where it may be sizable.  

Following Ref.\ \ci{passek}, we assume that mixing is completely 
embedded in the decay constants and use particle-independent
distribution amplitudes $\phi_{Pi}\equiv\phi_{i}$, where $P=\eta,
\eta^\prime$. As remarked above, the $P=\eta,\eta^\prime$ amplitudes  
are expressed in terms from the contributions of their flavour-octet and 
flavour-singlet $q \bar{q}$ Fock components as well as from the $gg$
one. The former are given by \req{eq:LOamp} with $f_P \to f_P^i$, 
$\phi_P \to \phi_i$, and ${\cal C}^{a {b}}_{P} \to {\cal C}^{a
  {b}}_{i}$, where $i=8,1$. The numerical values of the flavour factors 
are 
\be
\begin{array}{l}
\displaystyle
{\cal C}^{u {u}}_{8} = {\cal C}^{d {d}}_{8} 
=\frac{1}{\sqrt{6}} \, ,
\qquad {\cal C}^{s {s}}_{8} 
=-\frac{2}{\sqrt{6}} \, ,
\\[0.2cm] \displaystyle
{\cal C}^{u u}_{1} = {\cal C}^{d {d}}_{1} 
= {\cal C}^{s {s}}_{1} 
=\frac{1}{\sqrt{3}}
\, ,
\end{array}
\label{eq:Cabeta}
\ee
and ${\cal C}^{a {b}}_{i}=0$ for other flavour combinations. 
\begin{figure}
\centerline{
\includegraphics[width=12cm]{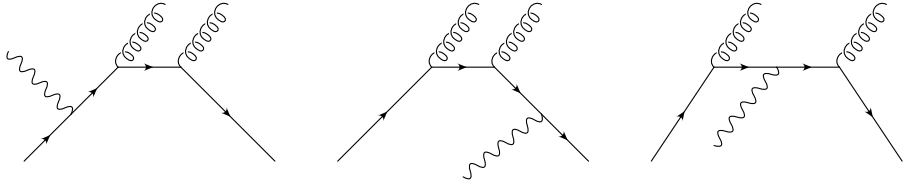}
}
\caption{LO Feynman diagrams that contribute
to the subprocess amplitude $\gamma q_a \to (gg) q_a$.
Three more graphs are obtained by crossing the gluon lines.}
\label{f:LOgg}
\end{figure}
The amplitudes $\gamma q_a \to (gg) q_a$ can easily be computed from
the Feynman diagrams displayed in Fig. \ref{f:LOgg}. Using the
normalisations of the two-gluon distribution amplitude and the
associated one of the two-gluon twist-2 projector as proposed in Ref.\
\cite{passek}, one finds the
following result for the invariant function $C_{\,2}^P$: 
\ba
C^{P(aa)}_{\,2} \Big|_{twist-2} & = &
        4\pi\als(\mu_R^2)\,\frac{C_F}{N_C}\, \frac{e_a }{\uh \sh}\, 
\Big\{ 
f_{P}^8 \,{\cal C}^{a{a}}_{8} \,\langle 1/\tau \rangle_8  \nn\\[0.3em]
 & & +f_{P}^1\,{\cal C}^{a{a}}_{1}\, \langle 1/\tau \rangle_1 
-f_{P}^1 \, \frac{1}{\sqrt{n_f}} \, \langle 1/\tau^2 \rangle_g
                      \Big\}\,, \nn\\[0.3em] 
C^{P(aa)}_{\,1} \Big|_{twist-2} & = & C^{P(aa)}_{\,3} \Big|_{twist-2} 
= C^{P(aa)}_{\,4} \Big|_{twist-2}=0\,,
\label{eq:A2tw2eta}
\ea
with $n_f=3$ being the numbers of quark flavours. The invariant
function  $C^{P(aa)}_{\,3}$ is zero to leading-twist accuracy since
the distribution amplitudes $\phi_{8,1}$ are symmetric under the
replacement $\tau\leftrightarrow 1-\tau$ and the gluon contribution
is zero even at parton level. The moment of the two-gluon
distribution amplitude, which is antisymmetric under the replacement 
$\tau\leftrightarrow 1-\tau$ and which mixes under evolution with 
$\phi_1$, is defined by
\be
\langle 1/\tau^2 \rangle_g = \int_0^1 d\tau \frac{\phi_g(\tau)}{\tau^2}\,. 
\ee

In \ci{passek}, the twist-2 gluon distribution $\phi_g$ amplitude has been
estimated from a NLO analysis of the $\gamma \to \eta, \eta^\prime$ transition
form factors. The Gegenbauer series of the  distribution
amplitudes have been truncated at $n=2$; for the only non-zero
Gegenbauer coefficient, a value of 
$B^{g}_2 \equiv B^{g}_2(\mu_0^2)=9\pm 12$ has been found
at a scale of $\mu_0^2=1\gev^2$.
This coefficient is connected, through evolution, with 
the Gegenbauer coefficient $B^{1}_2 \equiv B^{1}_2(\mu_0^2)=-0.08\pm 0.04$
appearing in the expansion of $\phi_1$. 
It is to be stressed that the evolutional effects play an important
role for the distribution amplitudes that differ from the asymptotic form
($\phi_1=\phi_8=6 \tau (1 - \tau)$, $\phi_g=0$).
Recently, a combined analysis of the transition form
factors and the inclusive $\Upsilon (1 S) \to \eta^\prime X$ decay 
lead to  more severe restrictions
of the Gegenbauer coefficients \ci{ali}, 
namely, $B^{(g)}_2 \approx 8\pm 5$ and $B^{1}_2 \approx -0.07\pm 0.03$
at $\mu_0^2=1\gev^2$. In any case, the two-gluon contribution may be
sizable and could enhance the cross section for $\eta^\prime$
photoproduction substantially.

\end{appendix}
    

\end{document}